\documentclass[aps,pre,onecolumn,showpacs,amsmath,amssymb,longbibliography,notitlepage,11pt,tightenlines,superscriptaddress]{revtex4-1} 
\usepackage[dvipsnames]{xcolor}
\usepackage{floatrow}
\usepackage{physics}
\usepackage{graphicx}
\definecolor{darkblue}{rgb}{0,0,0.6}
\definecolor{darkred}{rgb}{0.6,0,0}
\usepackage[colorlinks=true,urlcolor=darkblue,citecolor=darkblue, linkcolor=darkred,hyperfootnotes=false]{hyperref}

\newcommand{\Jij}{J_{ij}}

\begin{document}

\title{Mechanical hysterons with tunable interactions of general sign}

\author{Joseph D. Paulsen}
\email{paulse8@stolaf.edu}
\affiliation{Department of Physics, St.~Olaf College, Northfield, MN 55057}
\affiliation{Department of Physics and BioInspired Institute, Syracuse University, Syracuse, NY 13244}

\begin{abstract}
Hysterons are elementary units of hysteresis that underlie many complex behaviors of non-equilibrium matter. 
Because models of interacting hysterons can describe disordered matter, this suggests that artificial systems could respond to mechanical inputs in precise and targeted ways. 
Specifying the properties of hysterons and their interactions could thus be a general method for realizing arbitrary non-equilibrium behaviors. 
Elastic structures including slender beams, creased sheets, and shells are clear candidates for artificial hysterons, but complete control of their interactions has seemed impractical or impossible. 
Here we report a mechanical hysteron composed of rigid bars and linear springs, which has controllable properties and tunable interactions of general sign that can be reciprocal or non-reciprocal. 
We derive a mapping from the system parameters to the hysteron properties, and we show how collective behaviors of multiple hysterons can be targeted by adjusting geometric parameters on the fly. 
By transforming an abstract hysteron model into a physical design platform, our work demonstrates a route toward designed materials that can sense, compute, and respond to their mechanical environment. 
\end{abstract}

\maketitle

\bigskip

\medskip
\textbf{Introduction}
\medskip

When an amorphous solid, a disordered magnet, or a crumpled sheet is subjected to an oscillating global drive, it navigates a complex pathway through a multitude of locally-stable states \cite{Paulsen24}. 
A powerful approach for understanding these systems has been to focus on the localized hysteretic rearrangements---hysterons---that arise within them \cite{Preisach35,Falk11,Keim14,Mungan19,Mungan19b,Terzi20,Keim20,Keim22,Shohat23,Muhaxheri24,Martinez-Calvo24}. 
In these abstract hysteron models, each hysteron $i$ possesses a pair of thresholds $\gamma_i^+$ and $\gamma_i^-$ at which it switches between its two possible states, $s_i = \pm1$. 
More general models incorporate interactions between hysterons \cite{Keim21,Lindeman21,Hecke21,Szulc22}, so that the switching thresholds $\gamma_i^\pm$ depart from their ``bare'' values $\gamma_{i,0}^\pm$ as follows: 
\begin{equation}\label{eq:model}
\gamma_i^\pm(S) = \gamma_{i,0}^\pm - \sum_{j \neq i} \Jij s_j \ ,
\end{equation}
where $S$ is a microstate that specifies the $\{s_j\}_{j \neq i}$ 
and the $\Jij$ encode cooperative ($\Jij>0$) or frustrated ($\Jij<0$) interactions that may be non-reciprocal ($\Jij \neq J_{ji}$). 

In addition to offering a foothold for understanding disordered media, Eq.~\ref{eq:model} suggests a design principle for a new class of synthetic matter. 
If the thresholds and interactions can be designed, materials could be created with tailored responses to different mechanical environments. 
As the states $s_i$ are readily interpreted as bits, such materials are endowed with digital memory and the ability to perform computations \cite{Yasuda21}. 
Several mechanical metamaterials have been developed to this end, including origami bellows \cite{Jules22}, corrugated sheets \cite{Bense21}, buckled beams \cite{Sirote-Katz24,Liu24}, and biholar sheets \cite{Ding22,El-Elmi24}. 
Yet, even in this realm where one has complete control over the material architecture, no system has been able to realize the full generality of Eq.~\ref{eq:model}. 

Here, we translate this abstract model into a physical design platform. 
We show how mechanical hysterons built from rigid bars and linear springs can realize the full generality of Eq.~\ref{eq:model} by way of strong, tunable interactions that are ferromagnetic-like or anti-ferromagnetic-like. 
We then target three specific computations on sequences of mechanical inputs and build the corresponding systems. 
Our results provide a replicable, general platform for processing mechanical information \textit{in materia}. 

\medskip
\textbf{Results}
\medskip

\begin{figure}[p!]
\includegraphics[width=0.5\textwidth]{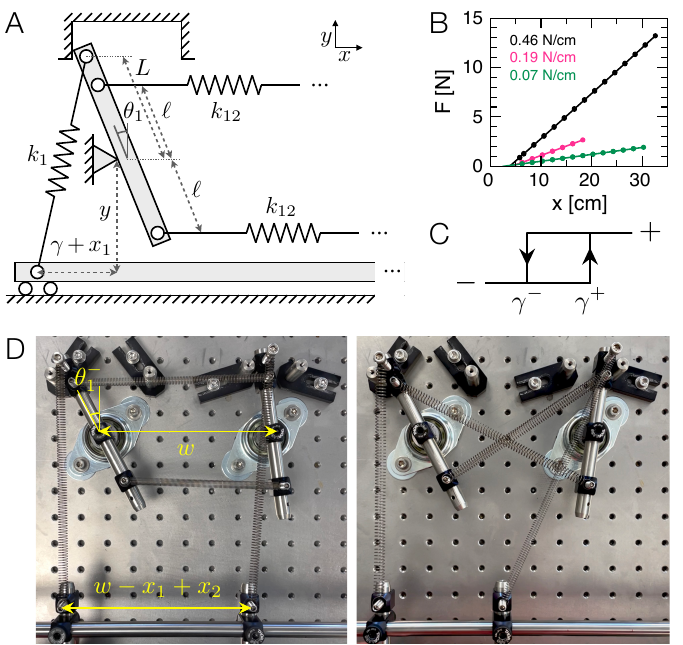}
\caption{
\textbf{Coupled mechanical hysterons with tunable interactions.}
A. Each hysteron is a rigid bar that may freely rotate between two hard boundaries, under the influence of a driving spring that attaches to a rod applying a horizontal quasistatic global drive, $\gamma$. 
Hysterons may be coupled together with additional springs. 
B. Typical force-versus-displacement data for springs used in the experiment. 
Solid lines are linear fits used to extract the stiffness $k$ and rest length, $x_0$. 
C. Basic behavior of a hysteron.  
The thresholds, $\gamma^-$, $\gamma^+$, depend on the geometric properties of the system as well as the state of the other hysterons when interactions are present. 
D. Experimental realizations with cooperative (left) and frustrated interactions (right). 
}
\label{fig:1}
\end{figure}

\textbf{Hysteron design.}
Figure~\ref{fig:1}A illustrates the basic bistable unit in our system: a rigid bar that is free to rotate about a central pivot. 
The bar is corralled between two posts that restrict its angle to an interval, $[\theta^-,\theta^+]$, with $\theta=0^\circ$ along the $y$-axis. 
A spring of stiffness $k_1$ and rest length $x_0$ attaches to the bar a distance $L$ from the pivot; the other end of this spring attaches to a long steel rod that translates freely along its axis via two linear bearings. 
The $x$-position of this rod, $\gamma$, serves as the global mechanical drive. 
All springs used in the experiments behave linearly, which we verify by measuring their force-versus-displacement curves when pulling up on a known mass resting on an electronic balance (Fig.~\ref{fig:1}B). 
We analyze these data to measure $k$ and $x_0$ for each spring used in the experiment. 

We first consider the stability of a single rotor. 
Starting with the driving rod on the far left and gradually moving it to the right, the rotor begins in the ($-$) state where $\theta=\theta^-$, but becomes unstable and jumps to the ($+$) state ($\theta=\theta^+$) when the spring crosses the pivot point of the rod. 
This transition occurs when the $x$-position of the driving rod, $\gamma$, exceeds a threshold $\gamma^+ = -y\tan \theta^-$, where $y$ is the fixed distance from the driving rod to the rotor pivots. 
Likewise, the rotor flips to $(-)$ when $\gamma < \gamma^- = -y\tan \theta^+$. 
The rotor is thus a hysteron: it is monostable outside the interval $[\gamma^-,\gamma^+]$ and it is bistable within (Fig.~\ref{fig:1}C). 
We may move this interval of stability by shifting the position $x_i$ where hysteron $i$ couples to the driving rod (where $x_i=0$ corresponds to no shift). 
Supplementary Movie 1 shows these basic behaviors for a pair of uncoupled rotors. 

Next, we create interactions by coupling rotors $i$ and $j$ together with additional springs of stiffness $k_{ij}$ and rest length $x_0$ mounted a distance $\ell$ from the pivot (Fig.~\ref{fig:1}A). 
Uncrossed coupling springs create cooperative (ferromagnetic-like) interactions that encourage two hysterons to occupy the same state, forming the composite states $--$ or $++$ more readily. 
Crossed springs create frustrated (antiferromagnetic-like) interactions that encourage them to occupy the states $+-$ or $-+$ (Fig.~\ref{fig:1}D). 

These interactions may be understood by torque balance. 
Suppose two rotors are held fixed at $\theta_1 = \theta_2 = 0^\circ$. 
Coupling the rotors with two identical springs leads to zero net torque on rotor $1$, as the torques from the two springs cancel whether they are crossed or uncrossed. 
Moving rotor $2$ into the $+$ state increases the distance spanned by one coupling spring and decreases the distance spanned by the other. 
Since these springs attach to rotor $1$ on different sides of the pivot, this change in the torque adds up: two positive numbers for uncrossed springs (pushing rotor 1 towards the $+$ state) and two negative values for crossed springs (pushing rotor 1 towards the $-$ state). 
Through these interactions, the switching thresholds for one hysteron thus become dependent on the state of the other, giving rise to an expanded set of thresholds. 
The ensuing complexity and rapid tunability is demonstrated in Supplementary Movie 2, where we demonstrate a multitude of transitions for two cooperatively-coupled hysterons.

\begin{figure}[p!]
\includegraphics[width=0.43\textwidth]{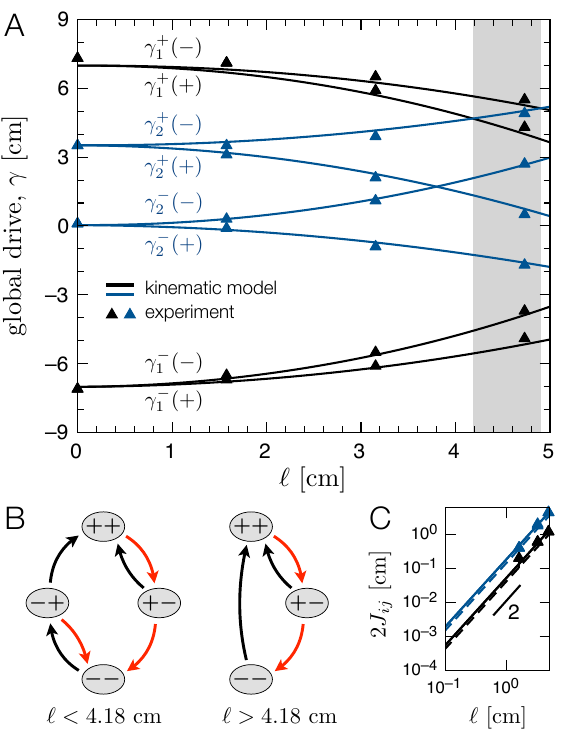}
\caption{
\textbf{Switching thresholds for an interacting hysteron pair.}
A. Using the configuration shown in the left panel of Fig.~\ref{fig:1}D, we measure the eight switching thresholds as the mounting position of the coupling springs, $\ell$, is varied. 
Experimental uncertainty is $\pm 0.1$ cm, which is approximately the size of the symbols. 
The experiments are in good agreement with the theory. 
Geometric parameters: $y = 9.4$ cm, $L = 6.3$ cm, $w = 15.2$ cm, $x_1 = 0$ cm, $x_2 = -1.9$ cm, $[\theta_1^-,\theta_1^+] = [-36.6^\circ, 36.7^\circ]$, and $[\theta_2^-,\theta_2^+] = [-9.8^\circ,11.2^\circ]$. 
Driving springs: $k_1 = k_2 = 0.19$ N/cm, rest lengths $3.9$ and $4.2$ cm. Coupling springs: $k_{12} = 0.07$ N/cm, rest lengths $2.4$ and $2.6$ cm. 
B. Transition graphs showing behavior for $\ell < 4.18$ cm and $4.18 < \ell < 4.90$ cm. 
When $\ell$ surpasses $4.90$ cm a more subtle change in behavior occurs; the avalanche from $--$ to $++$ passes through the intermediate state $+-$, as now $\gamma_1^+(+) < \gamma_2^+(-)$. 
C. Splitting between pairs of switching thresholds, $2\Jij = \gamma_i^\pm(-) - \gamma_i^\pm(+)$, which measures the strength of interactions between hysterons. Triangles: Experiments. Solid lines: Kinematic model. Dashed lines: Eq.~\ref{eq:Jij} for rotors confined to small angles. 
}
\label{fig:2}
\end{figure}

\textbf{Tunable interactions.}
The position where the coupling springs attach, $\ell$, offers a convenient way to tune the interaction between the hysterons continuously. 
To demonstrate this control, we perform experiments on two hysterons with a cooperative interaction ($\Jij>0$), where we fix all other system parameters and vary $\ell$. 
We measure the switching thresholds by holding one rotor fixed in the $(+)$ or $(-)$ state and quasistatically moving the driving rod until the other hysteron undergoes a transition. 
For example, $\gamma^+_1(s_2)$ denotes the measured threshold for hysteron 1 to switch to $(+)$ when hysteron 2 is in state $s_2$. 
Figure~\ref{fig:2}A shows the measured values of the eight transitions for this system at four values of $\ell$. 

When $\ell=0$, the coupling springs have no lever arm to exert a torque on either rotor. 
Here there are only four distinct thresholds due to the degeneracies $\gamma_i^\pm(+) = \gamma_i^\pm(-)$. 
The particular order of the thresholds, $\gamma^-_1 < \gamma^-_2 < \gamma^+_2 < \gamma^+_1$, gives rise to a set of stable states and transitions that we may represent in a transition graph \cite{Mungan19,Bense21} -- a directed graph that shows the transitions out of each stable state under increasing or decreasing drive (Fig.~\ref{fig:2}B). 
For $\ell=1.6$ cm, the thresholds split into eight distinct values. 
Nevertheless, the same stable states and transitions remain at $\ell=1.6$ cm and at $\ell=3.2$ cm. 

Increasing $\ell$ to $4.7$ cm, two pairs of thresholds switch their ordering: $\gamma_2^+(+)$ falls below $\gamma_2^-(-)$, and $\gamma_1^+(+)$ falls below $\gamma_2^+(-)$. 
The first switch does not change the transition graph for the system, but the second does. 
Starting in $--$, the first transition upon increasing $\gamma$  is to $-+$ at $\gamma=\gamma_2^+(-)$. 
But this state is now unstable, as $\gamma_1^+(+) < \gamma_2^+(-)$, meaning that hysteron 1 is above threshold to flip to $(+)$ as soon as hysteron 2 enters the $(+)$ state. 
This ``vertical avalanche'' \cite{Hecke21} is denoted by the  arrow from $--$ to $++$ in the second transition graph in Fig.~\ref{fig:2}B.  

This tunability of the thresholds and interactions suggests that our mechanical hysterons should be able to exhibit a wide variety of behaviors. 
We confirm this notion in Supplementary Movies 2-5, where we demonstrate all possible transition graphs for two hysterons with cooperative interactions ($\Jij > 0$) and frustrated interactions ($\Jij < 0$) \cite{Hecke21}. 
This suite of behaviors includes a 2-bit analog-to-digital converter (Fig.~S1A, transition graph $ix$). 
We provide further descriptions of these experiments in the Supplementary Information and in Fig.~S1.

\textbf{Kinematic model.} 
We can use torque balance to connect the system parameters to the switching thresholds observed in the experiment. 
First, we assume the system is in a stable microstate $S$. 
It is then straightforward to write down the torque $\tau_{ij}(s_i,s_j)$ of rotor $j$ on rotor $i$ in terms of the properties and mounting positions of the coupling springs. 
Similarly, one can compute the torque $\tau_i(\gamma,s_i)$ on rotor $i$ by its driving spring. 
Exact expressions for $\tau_i(\gamma,s_i)$ and $\tau_{ij}(s_i,s_j)$, which take into account the finite rest length of the springs, are provided in the Supplementary Information. 
Hysteron $i$ will undergo a transition when the sum of the torques on it vanish: 
\begin{equation}\label{eq:kinematic}
0 = \tau_i(\gamma,s_i) + \sum_{j \neq i} \tau_{ij}(s_i,s_j) \ .
\end{equation}
The $\gamma$ that solve this equation are the switching thresholds, $\gamma_i^\pm (S)$, wherein hysteron $i$ switches out of the state $s_i = \mp 1$. 
In principle, there could be other stable equilibria between $\theta_i^-$ and $\theta_i^+$; in practice we avoid this by using sufficiently strong driving springs compared to any coupling springs that might disrupt this bistability. 
We use Eq.~\ref{eq:kinematic} to predict thresholds for the system in Fig.~\ref{fig:2}A and find good agreement with the experiments. 
Moreover, the model allows us to anticipate precisely where the qualitative behaviors of the system will change, e.g., the change in the transition graph at $\ell = 4.18$ cm (Fig.~\ref{fig:2}A,B).

\textbf{Linear interactions at arbitrary rotor angles.} 
Equation \ref{eq:model} posits that the hysteron interactions are pairwise and linear. 
We now show that our system exhibits such interactions when the rest length $x_0$ of each driving spring is much less than $L+y$. 
In this limit, the torque on rotor $i$ by its driving spring is: $\tau_i = L k_i \cos \theta_i (\gamma + x_i + y \tan \theta_i)$. 
We emphasize that this result is for arbitrary $\theta_i$ and $\gamma$ as detailed in the Supplementary Information. 
Plugging this expression into Eq.~\ref{eq:kinematic} and solving for $\gamma$ gives the transitions $\gamma^\pm_i(S)$ for rotor $i$ when the system is in microstate $S$. 
The interaction of hysteron $j$ on hysteron $i$ may then be calculated by the difference, $\gamma_i^\pm(S |_{s_j=-}) - \gamma_i^\pm(S |_{s_j=+}) = 2\Jij$, which yields:  
\begin{equation}\label{eq:Jijgeneral}
\Jij^\pm = \frac{1}{2 L k_i \cos \theta_i^\mp} (\tau_{ij}(\mp,+) - \tau_{ij}(\mp,-)) \ ,
\end{equation}
independent of the state of all other rotors. 
Such a result establishes that the interactions are pairwise and linear, even for arbitrary rotor angles and finite coupling spring rest lengths. 
This linearity may be attributed to two key features of our system: (i) the affine relationship between the drive $\gamma$ and its torque on a rotor, and (ii) the additivity of the torques from the coupling springs. 

Equation~\ref{eq:Jijgeneral} additionally says that there are two sets of interactions strengths: $\Jij^+$ describes the interactions for hysteron $i$ when $\theta_i = \theta_i^-$, whereas $\Jij^-$ is the relevant matrix when $\theta_i = \theta_i^+$. 
Such abstract hysteron models have been proposed recently \cite{Szulc22}, where $\Jij$ in Eq.~1 is replaced by $\Jij^\pm$. 
In Fig.~S2 we show that the simpler scenario, $\Jij^+ = \Jij^- = \Jij$, occurs when the stopping posts are positioned symmetrically for each rotor so that $\theta_i^+ = -\theta_i^-$ for each $i$. 

We can also calculate the bare thresholds, $\gamma_{i,0}^\pm$, in terms of the torques on rotor $i$. 
They are found from the relation, $ \gamma^\pm_i(S^+) + \gamma^\pm_i(S^-) = 2 \gamma_{i,0}^\pm$, where $S^\pm$ is the state with $s_j = \pm 1$ for all $j \neq i$. 
Using Eq.~\ref{eq:kinematic} and the expression for $\tau_i$ above yields:
\begin{equation}\label{eq:gamma0general}
\gamma_{i,0}^\pm = -x_i - y\tan \theta_i^\mp - \frac{1}{2 L k_i \cos \theta_i^\mp} \sum_{j \neq i} (\tau_{ij}(\mp,+) + \tau_{ij}(\mp,-)) \ .
\end{equation}
We emphasize that the sum in Eq.~\ref{eq:gamma0general} represents a shift of the bare threshold due to the presence of interactions. 
Physically, it arises because the torque exerted by rotor $j$ on rotor $i$, averaged over the two states $s_j = \pm 1$, may be non-zero. 
By contrast, the interactions $\Jij$ measure the positive and negative departures from this mean, which are then added on top of this bare threshold, following Eq.~\ref{eq:model}.

\textbf{Mapping to hysteron model at small rotor angles.} 
We can gain a deeper understanding of the interactions by studying a limit where the rotor angles are small, $|\theta_i^\pm | \ll 1$. 
To emphasize the core relationships in this mapping, we specialize to the case where the two springs connecting rotors $i$ and $j$ have identical spring constants $k_{ij}$ and rest lengths $x_0$. 
To simplify expressions, we consider coupling springs with rest lengths that are small compared to either the rotor spacing or the mounting distance on the rotor: $x_0 \ll w$ or $x_0 \ll \ell$. 
We show in the Supplementary Information that under these conditions, the torque of hysteron $j$ on $i$ has a simple form: $\tau_{ij} = 2 g_{ij} k_{ij} \ell^2 (\theta_j - \theta_i)$, where the geometric factor $g_{ij}$ is defined by: 
\begin{equation}\label{eq:gij}
g_{ij} = \begin{cases}
			1 & \text{parallel springs,}\\
            -1 & \text{crossed springs.}
		 \end{cases}
\end{equation}
Plugging this torque into Eqs.~\ref{eq:Jijgeneral} and \ref{eq:gamma0general} at small angles, we obtain: 
\begin{eqnarray}
\Jij &=& g_{ij} \frac{k_{ij}}{k_i} \frac{\ell^2}{L} ( \theta_j^+ - \theta_j^- ) \ , \label{eq:Jij} \\
\gamma_{i,0}^\pm &=& -x_i - y\tan \theta_i^\mp - \sum_{j \neq i} \frac{k_{ij}}{k_i} \frac{\ell^2}{L} g_{ij} (\theta_j^+ + \theta_j^- - 2\theta_i^\mp) \ , \label{eq:bare}
\end{eqnarray}
where we retained $y \tan \theta_i^\mp$ in place of $y \theta_i^\mp$ in Eq.~\ref{eq:bare} so that it gives the exact threshold in the absence of coupling. 

Figure~S3 shows that the small-$\theta$ model of Eqs.~\ref{eq:model}, \ref{eq:Jij}, and \ref{eq:bare} provides a good approximation to the exact kinematic model, Eq.~\ref{eq:kinematic}, for both cooperative and frustrated interactions. 
Moreover, Eqs.~\ref{eq:Jij} and \ref{eq:bare} provide important physical insights into our system. 
For instance, they tell us that only one interaction matrix is sufficient to describe the system at small rotor angles ($\Jij^+ = \Jij^- = \Jij$), even without the additional symmetry of the rotor angles assumed in Fig.~S2. 
They also identify an inherent non-reciprocity of the interactions that we explore in the next section. 
More valuable still, Eqs.~\ref{eq:Jij} and \ref{eq:bare} are a blueprint for designing the thresholds and interactions in our system at will---an enticing topic that comprises the remainder of this article.

\textbf{Non-reciprocal interactions.}
Returning first to Fig.~\ref{fig:2}A, a closer look reveals that those interactions are non-reciprocal: The splitting of the blue curves, $\gamma_2^\pm(-) - \gamma_2^\pm(+)$, is larger than the splitting of the black curves, $\gamma_1^\pm(-) - \gamma_1^\pm(+)$. 
The non-reciprocity of the interaction can be seen immediately via Eq.~\ref{eq:Jij}, where $\Jij$ depends on the angular interval $\theta_j^+ - \theta_j^-$ but does not depend on $\theta_i^\pm$. 
The larger splitting of $\gamma_2$ in Fig.~\ref{fig:2}A is thus due to the larger angular interval of hysteron 1. 
To test this result quantitatively, Fig.~\ref{fig:2}C compares Eq.~\ref{eq:Jij} with the value of $2\Jij = \gamma_i^\pm(-) - \gamma_i^\pm(+)$ from the full kinematic model and the experiments. 
The comparison shows that Eq.~\ref{eq:Jij} captures the observed non-reciprocity as well as the predicted $\ell^2$ scaling. 

Having established a physical picture for the switching thresholds and interactions for pairs of mechanical hysterons, the remainder of this article is devoted to targeting more complex behaviors with two, four, and many coupled hysterons.

\textbf{Hysteron latching.}
Holding a blueprint for frustrated non-reciprocal interactions allows us to target an exotic behavior termed ``latching''. 
This behavior was proposed by Lindeman \textit{et al.}~\cite{Lindeman25} to explain how amorphous solids form multiple memories of asymmetric driving. 
Here we demonstrate it with our mechanical hysterons in Fig.~\ref{fig:3}A and Supplementary Movie 6. 

Starting in the $--$ state at $\gamma=0$, we drive the system to $\gamma=10$ cm, which puts the system in the $+-$ state. 
This state remains stable as $\gamma$ returns back to $0$; hysteron 1 is now ``latched'' in the $(+)$ state. 
A larger driving amplitude to $\gamma=32$ cm and back to $0$ releases hysteron 1 back to $(-)$ via a sequence of intermediate states: $+- \rightarrow ++ \rightarrow -+ \rightarrow --$. 

Latching necessarily violates return-point memory \cite{Barker83,Sethna93,Keim19}, which would stipulate that the same state would recur each time the driving returns to its common endpoint, $\gamma=0$. 
Indeed, Lindeman \textit{et al.}~\cite{Lindeman25} showed that latching requires a strong, frustrated, non-reciprocal interaction: $J_{12} < J_{21} < 0$. 
Figure~\ref{fig:3}B shows that our setup satisfies this requirement. 
We obtain this frustrated non-reciprocal  interaction by confining hysteron 1 to a smaller angular interval than hysteron 2 and coupling them with crossed springs. 
The full requirements for latching, and how we target them using our mechanical hysterons, is provided in the Supplementary Information.

\textbf{Counting driving cycles.}
Frustrated interactions enable other computational behaviors, such as counting the number of mechanical driving cycles that have been applied. 
Our design uses an antiferromagnetic chain of hysterons with uniform frustrated interactions and uniform hysteresis, $\gamma^+_i - \gamma^-_i$. 
We alternately raise and lower both of the bare switching thresholds for each hysteron (Fig.~\ref{fig:4}A). 
This staggering breaks the degeneracy of the two ground states so that $+-$$+-$$...$ has a higher energy than $-+$$-+$$...$. 
Consider now a stable state with one ``domain wall'' in the bulk where the two ground states meet. 
Both hysterons at the domain wall experience two interactions that cancel, lowering their barrier to change state via the global drive. 
This grows the $-+$$-+$$...$ domain by one site for each half-cycle of driving. 
The location of the domain wall thus encodes the number of driving cycles; $2n$ hysterons can record this number up to $n$. 

This design is easily translated to our mechanical system by forming a chain of rotors that are coupled to their nearest neighbors with crossed springs and attaching them to the driving rod with $x_i$ that alternate between negative and positive values (Fig.~\ref{fig:4}B). 
In particular, we set $x_{2m-1} = -L\sin(\theta_0)$, $x_{2m} = L\sin(\theta_0)$, where $\theta_0 = \theta_i^+ = - \theta_i^-$, although other values can also work (see the Supplementary Information). 
Finally, we set $x_1 = -3L\sin(\theta_0)$ at the start of the chain to nucleate a domain wall there. 
Supplementary Movie 7 and Fig.~\ref{fig:4}B show the system evolution, starting from the higher-energy ground state. 
Once the domain wall reaches the end of the chain, the configuration is stable to further driving.

\textbf{Counting modulo 2.}
Figure~\ref{fig:5}A shows a system of four hysterons that, when driven at unit amplitude, produces a steady state that takes two cycles to traverse. 
It concludes even cycles in the state $--$$-+$, and it concludes odd cycles in the state $--$$+-$ (Fig.~\ref{fig:5}B). 
The system thus performs a basic computation: discriminating even versus odd numbers of driving cycles, \textit{i.e.}, counting modulo 2. 
Such a multiperiodic response was identified in simulated hysteron systems \cite{Keim21,Lindeman21,Szulc22}. 
Here we realize it in the lab. 

We start by translating the sign and relative strength of the interactions in Fig.~\ref{fig:5}A into a choice of springs and mounting points on the rotors. 
One pair of springs spans a longer distance from rotor 1 to rotor 3; we attach a length of steel wire to extend these springs while staying within their linear response. 
We then set $\theta_i^+ - \theta_i^-$ for each hysteron to reflect its hysteresis in the design. 
To ensure constant-amplitude driving, we use the linear bearings that hold the driving rod as a ``sandwich'' that sets the endpoints of the driving. 

Allowing ourselves to vary the value of $\ell$ for each pair of rotors, the mounting positions $x_i$, the angular intervals $\theta_i^+ - \theta_i^-$, the coupling positions $\ell$, and the two endpoints of the driving offer a total of 14 continuous degrees of freedom. 
We also allow ourselves to mount some of the springs within a pair at unequal $\ell$. 
To target the sequence of states in Fig.~\ref{fig:5}B, we drive the system to observe its behavior and then adjust one or more of these degrees of freedom to try to obtain more of the desired sequence. 
This iterative process eventually produced the configuration shown in Fig.~\ref{fig:5}C; its behavior under cyclic driving is shown in Supplementary Movie 8. 
To show the resulting steady state, we measure the $\gamma$ where each transition occurs, and we plot the sequence of states as a function of time in Fig.~\ref{fig:5}D. 
The sequence matches that of Fig.~\ref{fig:5}B, it takes two periods to traverse, and it repeats indefinitely, as desired.

\medskip
\textbf{Discussion}
\medskip

Recent work has demonstrated mechanical designs that can count modulo two \cite{ten-Wolde24}, count driving cycles up to $n$ \cite{Kwakernaak23}, or encode an analog mechanical input into a binary string \cite{Hyatt23}. 
Although those systems use bistable mechanical elements, each one relies on a different strategy to navigate the pathways between its states. 
In contrast, we have presented a general platform that can be reconfigured to perform all of the above functions, plus a latching behavior that was recently identified in hysteron simulations \cite{Lindeman25}. 
This platform could find use in the design of mechanisms that navigate through precise sequences of states, e.g., in robotics \cite{Kamp24}. 
More broadly, our work identifies a vast potential for designed materials that inherently engage with memory \cite{Keim19}, as their current state is a by-product of their driving history along with their overall design, which our platform now brings under the experimenter's control. 

One principle of our approach was to use nonlinearities sparingly, so that the behaviors of the system could be rationalized as simply as possible. 
Nevertheless some important questions remain: Schrecengost~\cite{Schrecengost25} showed that mixed interactions ($\Jij J_{ji} < 0$) can be ruled out in a certain region of parameter space, but a full treatment of this question requires further work. 
Other designs for interacting mechanical hysterons are also being proposed; Shohat and van Hecke \cite{Shohat25} use theory and simulations to explore two-dimensional networks of bilinear hysterons as a route to exotic pathways, including mixed interactions and multiperiodic cycles. 
Understanding what are the general constraints on $\gamma_{i,0}^\pm$ and $\Jij$ for any realizable structure or metamaterial is an important open question \cite{Baconnier25}. 

Designing the response of larger collections of mechanical hysterons presents another challenge \cite{Teunisse25,Muhaxheri25}. 
In Fig.~\ref{fig:5} we obtained a period-2 response by aiming for a set of target hysteron parameters that were originally found via a random search~\cite{Keim21}. 
Our approach for dialing in that precise set of couplings was to iteratively adjust the mounting positions of springs and the post angles to pick up more and more of the sought after multiperiodic cycle. 
This process is reminiscent of supervised learning~\cite{Stern23,Altman24}, suggesting that learning might be a viable approach for targeting transition graphs at large. 
Our system offers a wealth of possible learning degrees of freedom, as there are numerous geometric parameters that can be tuned continuously to affect the switching thresholds, including $x_i$, $\theta_i^\pm$, $\ell$, and $L$. 
Further work is needed to determine whether a suitable learning algorithm could be used to navigate between different transition graphs, or classes of transition graphs with a desired functionality.

\newpage
\medskip
\textbf{Methods}
\medskip
 
\textbf{Rotor hysteron construction.}
We construct each rotor hysteron from two $1/2$"-diameter optical posts attached in a T configuration, which we mount in a radial steel ball bearing affixed to an optical table. 
We remove the bearing grease to further lower their friction to rotation. 
We use steel extension springs to couple the rotors to one another and to a 1/2"-diameter steel shaft that is mounted in two linear bearings. 
This 72''-long rod translates freely along the horizontal direction and provides the global mechanical drive, $\gamma$. 
We define $\gamma$ as the horizontal displacement of this rod in centimeters, which we measure with a precision of $0.1$ cm via a ruler affixed under the rod. 
We define $x_i$, the mounting position of driving spring $i$ on the driving rod, such that $x_i=0$ corresponds to no shift when the driving rod is at $\gamma=0$. 

\textbf{Spring stiffnesses.}
We measure the spring stiffnesses from force-versus-displacement data that we obtain by anchoring the springs to a known mass on an electronic balance, and recording the force on the balance as the top of the spring is extended and then retracted (Fig.~\ref{fig:1}B). 
These measurements show that the springs behave linearly over the range of extensions used in the experiments. 

\textbf{Abstract hysteron simulations.}
We verified the period-2 behavior of Fig.~\ref{fig:5}A using the open-source \texttt{hysteron} software package \cite{code}.

\medskip

\medskip
\textbf{Data and Code Availability}
\medskip
 
The data and custom code that support the findings of this article are openly available in the \textit{figshare} repository~\cite{PaulsenData25} under a CC BY 4.0 license.

\medskip
\textbf{Acknowledgements}
\medskip
 
I thank Nathan Keim for suggesting how to apply a global drive, for providing the parameters used in Fig.~\ref{fig:5}A, and for many fruitful discussions. 
I am grateful to Zachariah Schrecengost for identifying how the interactions could be treated at arbitrary rotor angles. 
I thank Vidyesh Anisetti, Eadin Block, Christian Santangelo, and Jennifer Schwarz for contributing to early hysteron designs, and I thank Martin van Hecke, Nathan Keim, Chloe Lindeman, Gentian Muhaxheri, and Dor Shohat for comments on the manuscript. 
This research was supported by Syracuse University research subsidy funds and St.~Olaf College research subsidy funds. 
The work was conceived at the Aspen Center for Physics, which is supported by National Science Foundation grant PHY-2210452.

\medskip
\textbf{References}
\medskip

%

\newpage

\begin{figure}[p!]
\includegraphics[width=0.5\textwidth]{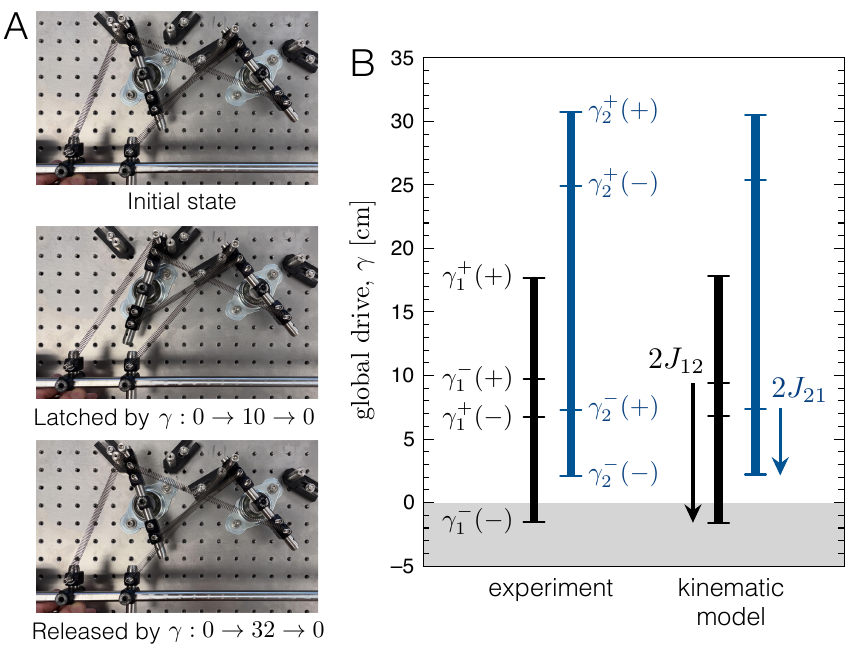}
\caption{
\textbf{Using a frustrated non-reciprocal interaction to build a hysteron latch.}
A. Coupled mechanical hysterons that are stable in the $+-$ and $--$ states at $\gamma=0$. 
Driving up to $\gamma=10$ cm and back to $0$ sets the system in $+-$. 
Driving up to $\gamma=32$ cm and back to $0$ resets the system to $--$. 
Geometric parameters: $y = 9.4$ cm, $L = 6.3$ cm, $\ell=4.7$ cm, $w = 15.2$ cm, $x_1 = -8.1$ cm, $x_2 = -16.4$ cm, $\theta_1^\pm = \pm18.3^\circ$, and $\theta_2^\pm = \pm38.8^\circ$. 
All springs have $k = 0.19$ N/cm and $x_0 = 4.1$ cm. 
B. Switching thresholds in the experiment and the model. 
Arrows identify the (negative) values of $2J_{12} = \gamma_1^-(-) - \gamma_1^-(+)$ and $2J_{21} = \gamma_2^-(-) - \gamma_2^-(+)$, highlighting the frustrated non-reciprocal interaction at the core of the behavior. 
Experimental uncertainty is $\pm 0.1$ cm (not plotted). 
}
\label{fig:3}
\end{figure}

\begin{figure}[p!]
\includegraphics[width=0.5\textwidth]{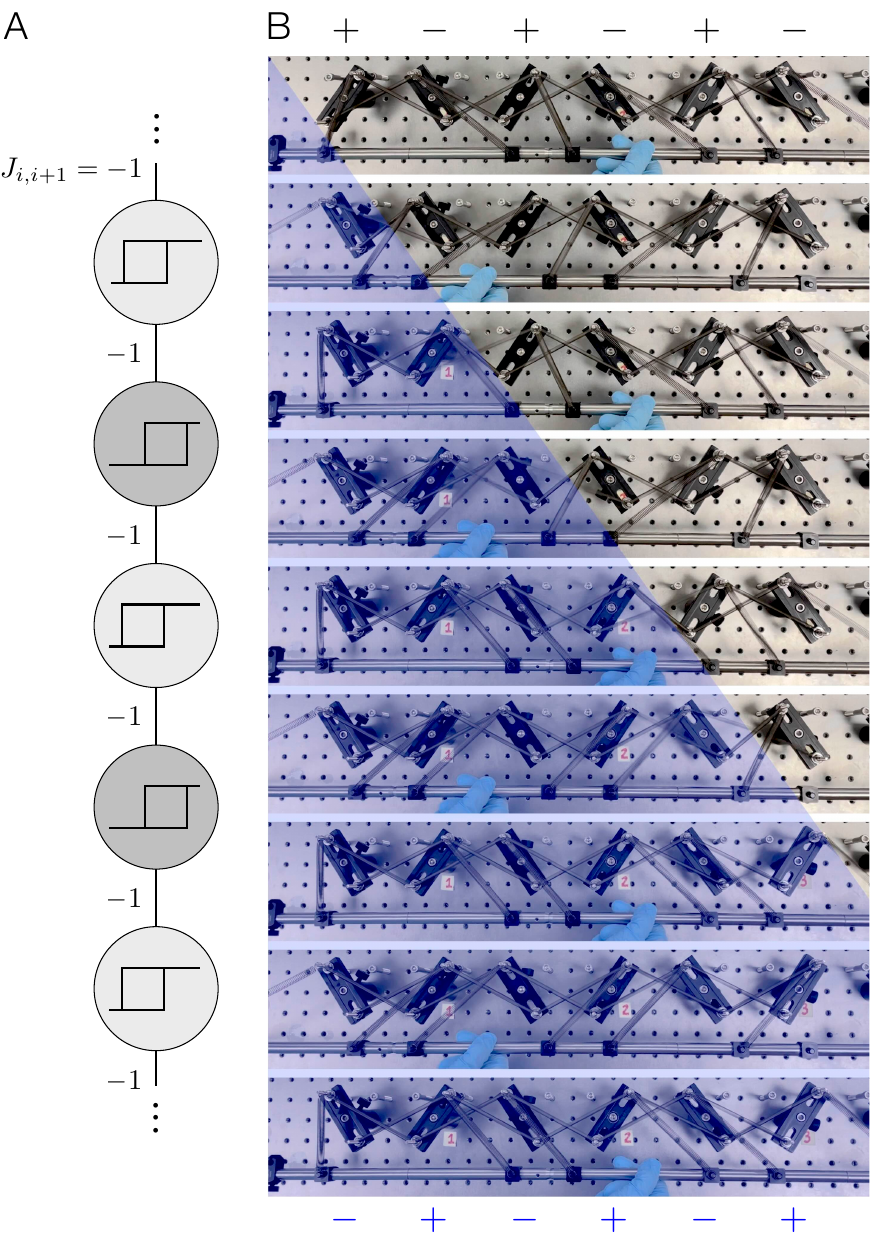}
\caption{
\textbf{Counting driving cycles with a chain of mechanical hysterons.}
A. One-dimensional chain of hysterons with uniform antiferromagnetic nearest-neighbor interactions. 
All hysterons have the same hysteresis, $\gamma^+_i - \gamma^-_i$, but the thresholds are shifted positively/negatively for odd/even $i$ (dark/light gray in schematic). 
B. Time series showing a domain wall ratcheting down the chain, irreversibly flipping one hysteron each half cycle. 
The system is initialized in the state $+-$$+-$$+-$ and a cycle of driving follows $\gamma: \gamma_0 \rightarrow -\gamma_0 \rightarrow \gamma_0$. 
The growing $-+$$-+$$...$ phase is highlighted in blue. 
For ease of construction, the radial bearings are replaced by optical posts in post holders lubricated with silicone oil. 
Coupling springs: $k=0.17$ N/cm, $x_0 = 4.0$ cm. 
Driving springs: $k=0.19$ N/cm, $x_0 = 4.1$ cm. 
}
\label{fig:4}
\end{figure}

\begin{figure}[p!]
\includegraphics[width=1.0\textwidth]{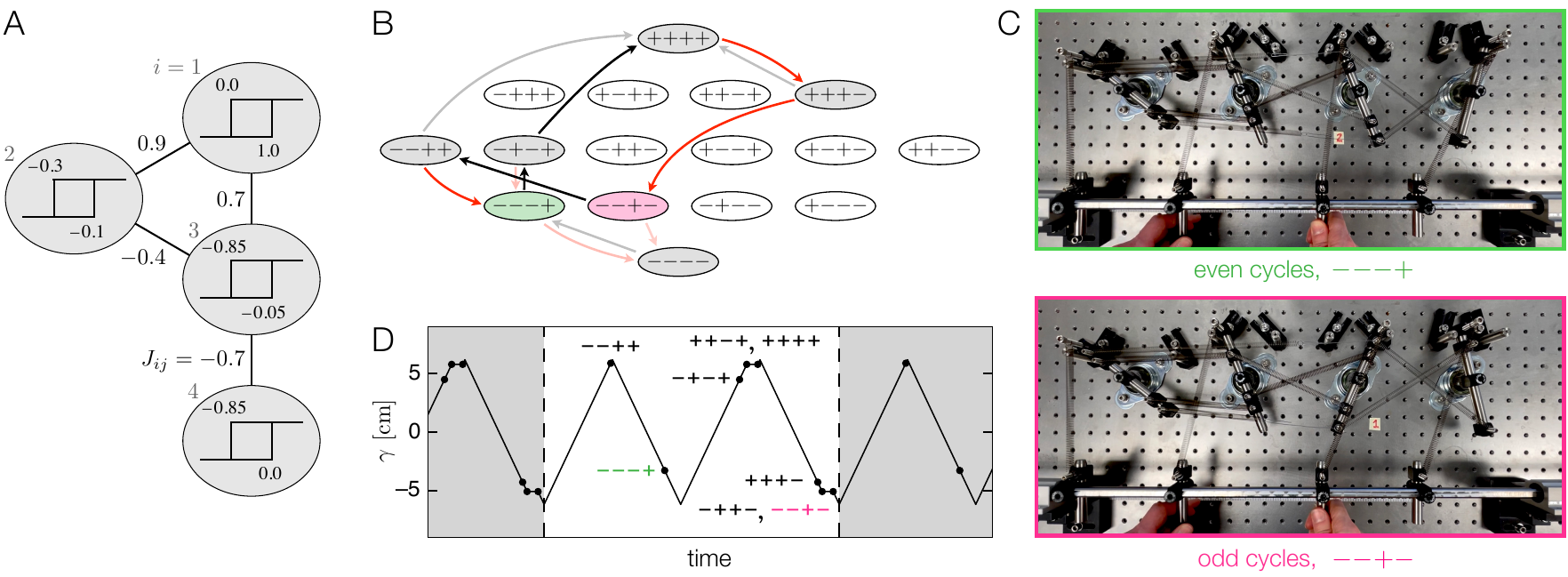}
\caption{
\textbf{Counting modulo 2 with four mechanical hysterons.}
A. Thresholds and interactions that yield a period-2 limit cycle for unit-amplitude cyclic drive. 
B. Corresponding pathway, where we have colored the two states at the minimum of the driving. 
Lighter arrows show all other transitions available to the system under arbitrary driving and are not used in the period-2 cycle. 
States without arrows to them are unstable at all $\gamma$. 
C. Realizing this pathway with mechanical hysterons. 
The rotors from left to right correspond to the hysterons from top to bottom in A. 
The springs connecting rotors 1 and 3 have $k=0.07$ N/cm, and their original $x_0 = 2.5$ cm are extended by steel wire. 
All other springs have $k=0.19$ N/cm and $x_0 = 4.1$ cm. 
D. States and transitions in the experiment, which achieves the targeted response. 
Horizontal segments denote avalanches. 
}
\label{fig:5}
\end{figure}



\newpage
\clearpage

\renewcommand{\thefigure}{S\arabic{figure}}
\setcounter{figure}{0} 
\renewcommand{\theequation}{S\arabic{equation}}
\setcounter{equation}{0}

\centerline{\textbf{\large Supplementary Information for }} 
\centerline{\textbf{\large ``Mechanical hysterons with tunable interactions of general sign"}}
\medskip
\centerline{{ Joseph D. Paulsen}}

\section{Targeting specific transition graphs}
We start by demonstrating four distinct transition graphs in a sequence where we alter only the mounting position of driving spring 2, thereby changing only $x_2$. 
Supplementary Movie 2 shows this sequence of behaviors, which correspond to transition graphs $i$-$iv$ in Fig.~\ref{fig:4}A. 
The experimental parameters are tabulated in Fig.~\ref{fig:4}C. 

To understand how the system moves between these four transition graphs, we plug these parameters into the kinematic model, calculate the switching thresholds, and plot them in Fig.~\ref{fig:4}B. 
The switching thresholds of hysteron 1 are constant throughout this sequence, as they depend on the state of hysteron 2 but not its driving. 
The thresholds for hysteron 2 move in concert without changing their spacing, as changing $x_2$ shifts each threshold by $x_2$. 
Careful inspection of the switching thresholds in Fig.~\ref{fig:4}B shows that they indeed produce graphs $i$-$iv$ in Fig.~\ref{fig:4}A, faithful to the experiment. 

Next, we access transition graph $v$ by decreasing $\Delta \theta_1$, increasing $\Delta \theta_2$, and changing $x_2$ (Supplementary Movie 3). 
We access transition graph $vi$ by setting $\Delta \theta_1$ = $\Delta \theta_2$ and changing $x_2$ once again (Supplementary Movie 4). 
This exhausts all possible transition graphs for two hysterons with cooperative interactions \cite{Hecke21}. 

Three additional transition graphs can be obtained by using frustrated interactions \cite{Hecke21}; they are graphs $vii$-$ix$ in Fig.~\ref{fig:4}A. 
Supplementary Movie 5 shows how we can realize these graphs in the experiment while varying only $x_2$ to navigate between them. 
All three graphs feature at least one ``horizontal avalanches''. 
To understand how they occur, start in the $--$ state in graph $vii$. 
The corresponding thresholds in Fig.~\ref{fig:4}B show that upon increasing $\gamma$, hysteron 2 switches first, as $\gamma_2^+(-) < \gamma_1^+(-)$. 
Now in the $-+$ state, the next event is at $\gamma = \gamma_1^+(+)$ into $++$. 
But this state is not stable, as this $\gamma$ is below $\gamma_2^-(+)$, which brings the system immediately to $+-$. 
An analogous process underlies each of the $-+ \leftrightarrow +-$ avalanches in graphs $vii$-$ix$. 

Among these graphs, perhaps the most exotic is $ix$, which navigates the sequence: $--$, $-+$, $+-$, $++$ under unidirectional driving. 
This behavior constitutes a 2-bit analog-to-digital converter \cite{Hecke21}. 
We now describe the process used to target this transition graph, to illustrate how behaviors may be rationally designed in our system. 
First, we need a strong coupling between the global drive and hysteron 1, so that it is $(-)$ at small $\gamma$ and $(+)$ at large $\gamma$. 
We also need a strong frustrated interaction between hysterons 1 and 2 to navigate the $-+ \leftrightarrow +-$ avalanches. 
In particular, the interaction of hysteron 1 on hysteron 2 must open up a central interval, $\gamma_2^+(-) < \gamma < \gamma_2^-(+)$ where the state of hysteron 2 depends only on the current state of hysteron 1. 
Figure~\ref{fig:4}D illustrates this condition, and an examination of graphs $vii$-$ix$ in Fig.~\ref{fig:4} shows that they satisfy it. 
We target this condition by selecting a large $k_{12}$ and a sufficiently large $\theta_1^\pm$; we find that the same $\pm 38.8^\circ$ used in the other designs works here too. 
Finally, a weak coupling of hysteron 2 to the global drive enables a transition into the $++$ state at large $\gamma$ and into $--$ at small $\gamma$. 
These considerations lead to a tentative design for the spring constants: $k_1 \approx k_{12} \gg k_2$. 
Here we use $k_1 = k_{12} = 0.46$ N/cm and $k_2 = 0.07$ N/cm, which we found to suffice. 

Two additional transition graphs are possible for ``mixed'' interactions, where the signs of $\Jij$ and $J_{ji}$ differ \cite{Hecke21}. 
We do not see a way to realize such interactions in our system.

\section{Torque from the driving rod}
Here we derive the torque exerted by the driving rod on a hysteron. 
A driving spring of stiffness $k_i$ and rest length $x_0$ mounts to rotor $i$ at a position $L>0$ from the fulcrum of the rotor, with coordinates $\vec{r}_L = \{ L \sin \theta_i , L \cos \theta_i \}$. 
The other end of the driving spring connects to the driving rod at coordinates $\vec{r}_\text{bar} = \{ \gamma + x_i , -y \}$. 
The springs in the experiment obey Hooke's Law: $F = k_i (x - x_0)$, where $x = | \vec{r}_\text{bar} - \vec{r}_L |$. 
In vector form,
\begin{equation}
\vec{F} = k( \vec{r}_\text{bar} - \vec{r}_L ) \left( 1 - \frac{x_0}{ | \vec{r}_\text{bar} - \vec{r}_L | } \right) \ .
\end{equation}
The torque on rotor $i$ is $\tau_i = - \vec{r}_L \times \vec{F}$, where use a minus sign when defining the torque, so that clockwise rotation of the rotors corresponds to increasing $\theta$. 
Computing the cross product by components, we obtain the exact expression: 
\begin{equation}\label{eq:drive_exact}
\tau_i = L k_i \cos \theta_i ( \gamma + x_i + y \tan \theta_i ) \left( 1 - \frac{x_0}{\sqrt{ ( \gamma + x_i - L \sin \theta_i )^2 + (y + L \cos \theta_i)^2 }} \right) \ .
\end{equation}
We use Eq.~\ref{eq:drive_exact} when computing switching thresholds in the kinematic model. 

When the rest length of the driving spring, $x_0$, is small compared to $L+y$, the torque is approximately:
\begin{equation}\label{eq:drive_approx}
\tau_i = L k_i \cos \theta_i ( \gamma + x_i + y \tan \theta_i ) \ .
\end{equation}
We use Eq.~\ref{eq:drive_approx}  to derive Eqs.~3 and 4 in the main text and for the small-$\theta$ model.

\section{Torques between coupled hysterons}
Here we derive expressions for the torque exerted by one hysteron on another. 
We begin with the most general case: arbitrary angles for the rotors, finite rest lengths for the springs, and arbitrary mounting positions of the springs on the rotors (Eq.~\ref{eq:exact1}). 
We then expand this result to leading order in $\theta_1, \theta_2$ (Eq.~\ref{eq:approx1}). 
We then specialize this small-angle result to the ``paired springs" that are used exclusively in the main text. 
The result for the ``parallel'' configuration turns out to be independent of the rest length of the springs (Eq.~\ref{eq:parallel}), whereas the ``crossed'' configuration requires one more limit to reach a similarly simple expression (Eq.~\ref{eq:crossed}). 

\subsection{Single coupling spring}
We consider a single coupling spring of stiffness $k$ and rest length $x_0$ that mounts to rotors $1$ and $2$ at positions $\ell_1$ and $\ell_2$ from the fulcrums of the rotors, respectively, where the signs of $\ell_1$, $\ell_2$ indicate the side of the fulcrum where they attach. 
The coordinates of the mounting position on rotor $1$ are $\vec{r}_1 = \{\ell_1 \sin \theta_1 ,  \ell_1 \cos \theta_1 \}$, and on rotor $2$ they are $\vec{r}_2 = \{w+\ell_2 \sin \theta_2 ,  \ell_2 \cos \theta_2\}$. 
The springs in the experiment obey Hooke's Law: $F = k(x - x_0)$, where $x = |\vec{r}_2 - \vec{r}_1|$. 
In vector form,
\begin{equation}
\vec{F} = k(\vec{r}_2 - \vec{r}_1) \left( 1 - \frac{x_0}{ |\vec{r}_2 - \vec{r}_1| } \right) \ .
\end{equation}
The torque on rotor $1$ is $\tau = - \vec{r}_1 \times \vec{F}$, where use a minus sign when defining the torque, so that clockwise rotation of the rotors corresponds to increasing $\theta$. 
Computing the cross product by components, we obtain the exact expression: 
\begin{equation}\label{eq:exact1}
\tau = k \ell_1 ( w \cos \theta_1 + \ell_2 \sin (\theta_2 - \theta_1) ) \left( 1 - \frac{x_0}{\sqrt{ ( w + \ell_2 \sin \theta_2 - \ell_1 \sin \theta_1 )^2 + (\ell_2 \cos \theta_2 - \ell_1 \cos \theta_1 )^2 }} \right) \ .
\end{equation}
Expanding Eq.~\ref{eq:exact1} in $\theta_1$ and $\theta_2$, we obtain the torque on hysteron 1 to leading order in $\theta_1, \theta_2$:
\begin{equation}\label{eq:approx1}
\tau = k  \ell_1 ( w + \ell_2 (\theta_2 - \theta_1) ) \left( 1 - \frac{x_0}{L_0} \right) + k \ell_1 (\ell_2 \theta_2 - \ell_1 \theta_1) \frac{ w^2 x_0}{ L_0^3 } + ... \ ,
\end{equation}
where $L_0 = \sqrt{(\ell_2 - \ell_1)^2 + w^2}$ is the distance spanned by the spring when $\theta_1 = \theta_2 = 0$. 

Equations~\ref{eq:exact1} and \ref{eq:approx1} apply when rotor $2$ is to the right of rotor $1$. 
If rotor $2$ is on the left, sending $\ell_1 \rightarrow -\ell_1$ and $\ell_2 \rightarrow -\ell_2$ gives the relevant formulas, as this represents a $180^\circ$ rotation about the center of rotor $1$, which leaves both rotor angles and the coupling spring length unchanged. 


\subsection{Paired springs at small rotor angles}
Using Eq.~\ref{eq:approx1}, it is straightforward to write down the net torque $\tau_{12} = \tau + \tau'$ on rotor 1 for the general case where a second spring of stiffness $k'$ and rest length $x_0'$ is mounted at distinct positions $\ell_1'$ and $\ell_2'$. 
Pairing the springs as in Fig.~1D with spring constants $k = k'$ and $x_0 = x_0'$ greatly simplifies the form of the interaction. 
Mounting the springs in a ``parallel'' configuration, $\ell = \ell_1 = \ell_2 = -\ell_1' = -\ell_2'$, we find:
\begin{equation}\label{eq:parallel}
\tau_{12} = 2 k \ell^2 (\theta_2 - \theta_1) \ .
\end{equation}
Notably, the net torque does not depend on the finite rest length of the springs, $x_0 = x_0'$, or the distance between rotors, $w$, due to the symmetry of such paired coupling springs. 

Mounting instead in a ``crossed'' configuration, $\ell = \ell_1 = -\ell_2 = -\ell_1' = \ell_2'$, we obtain:
\begin{equation}
\tau_{12} = -2 k \ell^2 (\theta_2 - \theta_1) - 2 k \ell^2 \frac{ x_0 }{ \sqrt{4\ell^2 + w^2} } \left( 2\theta_1 - \frac{ 4\ell^2 }{ 4\ell^2 + w^2 } (\theta_2 + \theta_1) \right) \ .
\end{equation}
Specializing to either $x_0 \ll w$ or $x_0 \ll \ell$ simplifies this expression to:
\begin{equation}\label{eq:crossed}
\tau_{12} = -2 k \ell^2 (\theta_2 - \theta_1) \ , 
\end{equation}
which is equal and opposite to Eq.~\ref{eq:parallel}.




\section{Targeting a latching behavior}
In Fig.~3 of the main text we demonstrate a ``latching'' behavior, which had been introduced in the context of sheared amorphous solids in recent work by Lindeman \textit{et al.}~\cite{Lindeman25}. 
Here we describe how to rationally target this behavior using our mechanical hysterons. 
Lindeman \textit{et al.}~\cite{Lindeman25} showed how latching is tied to the ordering of the switching thresholds. 
They found that latching requires:
\begin{subequations}\label{eq:latch}
\begin{align}\label{eq:latcha}
& \gamma_1^-(-) < 0 < \gamma_2^-(-) < \gamma_2^-(+) < \gamma_1^-(+) \ , \\
& \gamma_1^+(-) < \gamma_2^+(-) \ . \label{eq:latchb}
\end{align}
\end{subequations}

We now show how we translate this ordering into a concrete design in our mechanical system. 
We start by noting that Eq.~\ref{eq:latcha} implies: $0 < \gamma_2^-(+) - \gamma_2^-(-) < \gamma_1^-(+) - \gamma_1^-(-)$, which we rewrite as: $0 < -J_{21} < -J_{12}$. 
We obtain this non-reciprocal frustrated interaction by confining hysteron 1 to a smaller angular interval than hysteron 2 and coupling them together with crossed springs following Eq.~6. 
Next, we note that adjusting $x_2$ moves all thresholds for hysteron 2 in concert without altering their spacing. 
We thus move $x_2$ until $\gamma_2^-(+) < \gamma_1^-(+)$ and $\gamma_1^-(-) < \gamma_2^-(-)$. 
Then, we fulfill $\gamma_1^-(-) < 0 < \gamma_2^-(-)$ by moving $x_1$ and $x_2$ together by equal amounts until those thresholds straddle $\gamma=0$. 
Finally, $\gamma_1^+(-) < \gamma_2^+(-)$ turns out to be satisfied already through the larger angular interval of hysteron 2. 


\section{Counting driving cycles}
Here we provide additional details regarding the accumulator design in the main text (Fig.~4). 
The basic idea of our design can be verified on a computer \cite{code} by applying cyclic driving of any fixed amplitude $\gamma_0$, setting thresholds $\gamma_{2m-1}^+ > -\gamma_{2m-1}^- = \gamma_0/2$ for the odd hysterons and $\gamma^\pm_{2m} = -\gamma^\mp_{2m-1}$ for the even hysterons (where $m=1,2,3...$), and setting nearest-neighbor interactions of strength $J_{i,i+1}=-\gamma_0$. 
The thresholds for hysteron 1 should be raised by $\gamma_0$ to allow a domain wall to nucleate there. 

More precisely, our accumulator design detects cycles of intermediate amplitude. 
Small-amplitude driving $\gamma(t)$ that is bounded by $\max(\gamma_i^-) < \gamma(t) < \min(\gamma_i^+)$ will not change the state of any hysteron, whereas large-amplitude driving that overpowers the frustrated interactions in the bulk can send the system to its final absorbing state. 
Intermediate-amplitude cycles advance the domain wall by two hysterons (one per half-cycle).


\newpage

\begin{figure*}[h]
\vspace{-0.0in}
\includegraphics[width=1.0\textwidth]{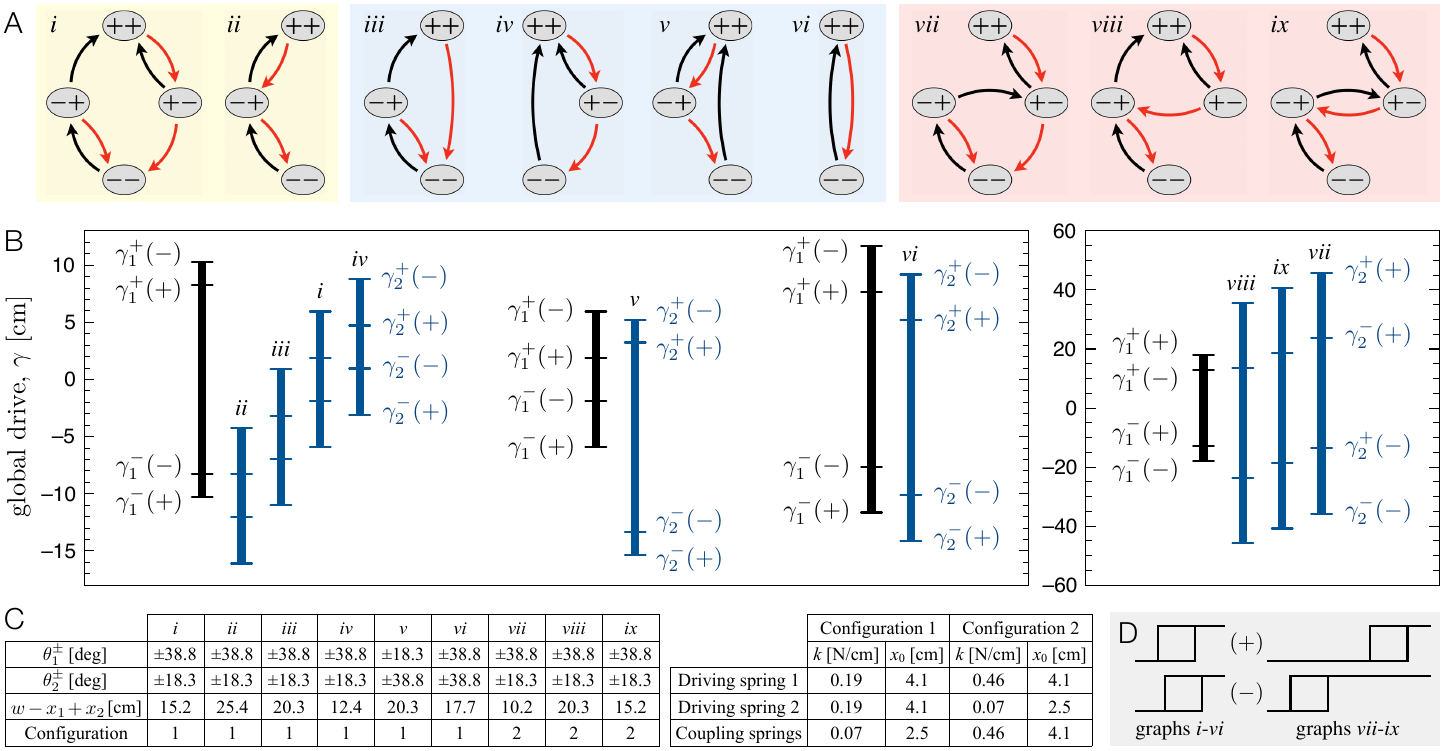}
\caption{
\textbf{Suite of behaviors for an interacting hysteron pair.}
A. Transition graphs obtained in the experiment, showing all states and transitions accessible from the initial state, $-$$-$ (Supplementary Movies 2-5). 
Graphs $i$ and $ii$ are ``Preisach'' graphs that do not require interactions \cite{Hecke21}, whereas graphs $iii$-$vi$ require cooperative interactions. To show how one can move between different transition graphs, we demonstrate them all using cooperative interactions, without changing the driving or coupling springs (``Configuration 1'' in C). 
Graphs $vii$-$ix$ require frustrated interactions; we obtain them with a second set of driving and coupling springs (``Configuration 2'' in C). 
B. Switching thresholds calculated from the model, using the experimental parameters. 
C. Experimental parameters. We fix $x_1=0$ cm, $y = 14.5$ cm, $L = 6.3$ cm, $\ell=4.7$ cm, and $w = 15.2$ cm. 
D. Schematic illustrating the role of interactions on hysteron 2 in graphs $i$-$ix$. Top row: hysteron 1 is $(+)$; bottom row: hysteron 1 is $(-)$. 
Smaller cooperative interactions are required to create graphs $i$-$vi$ in B, whereas larger frustrated interactions are needed for graphs $vii$-$ix$. 
}
\label{fig:4}
\end{figure*}

\newpage

\begin{figure*}[h]
\vspace{-0.0in}
\includegraphics[width=0.7\textwidth]{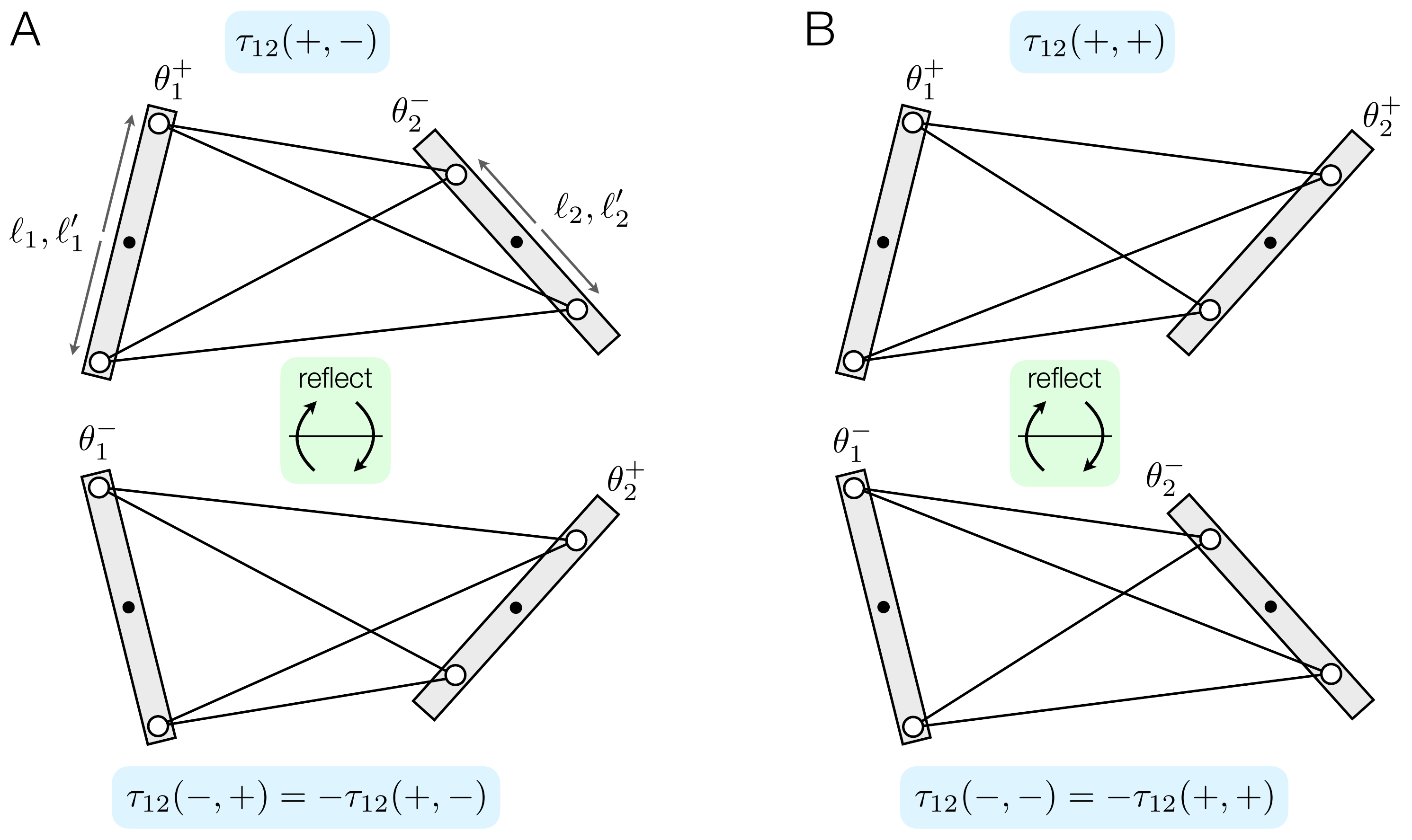}
\caption{
\textbf{Diagram showing equality of $\Jij^+$ and $J_{ij}^-$ when $\theta_i^+ = -\theta_i^-$ and $\ell_i = -\ell_i'$ for all $i$.}
A. Top schematic: Generic configuration with torque $\tau_{12}(+,-)$ on rotor $1$. 
Locations of both parallel and crossed springs are illustrated, but only one set is present at a time. 
If $\theta_i^+ = -\theta_i^-$ and $\ell_i = -\ell_i'$ for $i=1,2$, then reflecting this schematic about the horizontal is equivalent to flipping the states of both rotors, implying $\tau_{12}(-,+) = -\tau_{12}(+,-)$. 
B. Analogous pair of schematics showing $\tau_{12}(-,-) = -\tau_{12}(+,+)$ under the same conditions. 
Using Eq.~3 in the main text, we have $J_{12}^+ = (\tau_{12}(-,+) - \tau_{12}(-,-))/C^- = (\tau_{12}(+,+) - \tau_{12}(+,-))/C^+ = J_{12}^-$, where $C^\mp = 2 L k_i \cos \theta_i^\mp$. 
}
\label{fig:S2}
\end{figure*}

\newpage

\begin{figure}[h]
\vspace{-0.0in}
\includegraphics[width=0.5\textwidth]{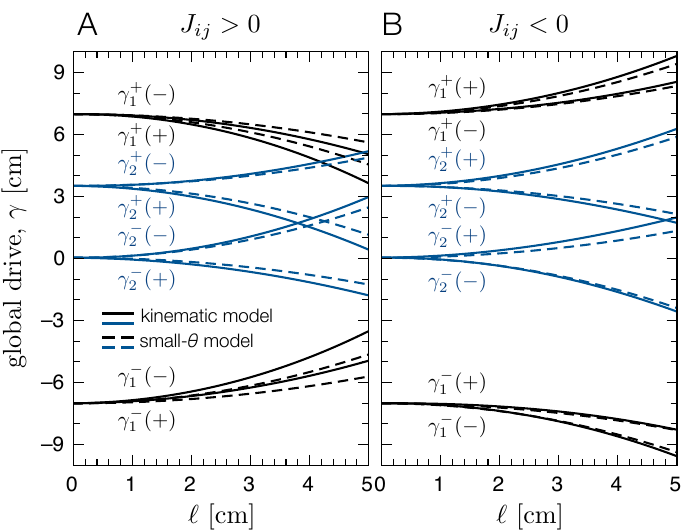}
\caption{
\textbf{Cooperative and frustrated interactions.}
A. We use the same system parameters as in Fig.~2 and calculate switching thresholds from the kinematic model (solid lines, Eq.~2) and the small-$\theta$ model (dashed lines, Eqs.~1, 6, and 7). 
The small-$\theta$ model captures the gross behavior of the system in spite of the large $\theta_i$ involved. 
B. Same analysis but with crossed springs, producing a frustrated interaction ($\Jij < 0$) so that now $\gamma_i^\pm (+) > \gamma_i^\pm (-)$. 
}
\label{fig:3}
\end{figure}

\end{document}